\documentclass[prd,aps,floats,twocolumn,nofootinbib,superscriptaddress]{revtex4-1}
\usepackage{slashed}
\usepackage{mathtools}
\usepackage{amsfonts}
\usepackage{amssymb}
\usepackage{epsfig}
\usepackage{caption}
\usepackage{subcaption}
\usepackage{rotating}
\usepackage{url}
\usepackage{times}
\usepackage{color}
\usepackage{bm}
\usepackage{xcolor,colortbl}
\usepackage{hyperref}
\usepackage[T1]{fontenc}
\usepackage{enumitem}
\usepackage{fancyhdr}
\usepackage{anyfontsize}
\usepackage{wasysym}
\usepackage{empheq}

\usepackage{simpler-wick}
\usepackage[compat=1.0.0]{tikz-feynman}
\newcommand\barparen[1]{\overset{\scriptscriptstyle{(-)}}{#1}}

\newcommand{\be}{\begin{equation}}
\newcommand{\ee}{\end{equation}}
\newcommand{\bea}{\begin{eqnarray}}
\newcommand{\eea}{\end{eqnarray}}

\begin{document}


\title{Fast Exact Algorithm for Neutrino Oscillation in Constant Matter Density}

\author{James Page}
\email{j.page@sussex.ac.uk}
\affiliation{Experimental Particle Physics Group, Department of Physics and Astronomy, Pevensey II Building, University of Sussex, Falmer, Brighton, BN1 9QH, United Kingdom}
\date{\today}

\begin{abstract}
A recently published method \cite{shafaq2020effect, bustamante2019nuoscprobexact} for solving the neutrino evolution equation with constant matter density is further refined and used to lay out an exact algorithm for computing oscillation probabilities, which is moderately faster than previous methods when looping through neutrinos of different energies. In particular, the three examples of $\barparen{\nu}_e$ survival, $\barparen{\nu}_\mu$ survival and $\barparen{\nu}_e$ appearance probabilities are written in terms of mixing angles, mass differences and matter electron density. A program based on this new method is found to be roughly twice as fast as, and in agreement with, the leading GLoBES package. Furthermore, the behaviour of all relevant effective parameters is sketched out in terms of a range of neutrino energies, or matter electron densities. For instance, the $\barparen{\nu}_e$ survival probability in constant matter density is found to have no dependence on the mixing angle $\theta_{23}$ or the CP-violating phase $\delta_{13}$.
\end{abstract}

\maketitle


\section{Introduction}

\subsection{The problem}

Neutrinos are produced in charged current (CC) interactions in pure flavour states $|\nu_{\alpha}\rangle, \; \alpha \in \{e, \mu, \tau \}$, which are composed of a superposition of the mass states $|\nu_{k}\rangle, \; k \in \{1, 2, 3\}$
\begin{equation}
    |\nu_{\alpha}\rangle = \sum_k U_{\alpha k}^* |\nu_{k}\rangle,
    \label{1a1}
\end{equation}
where it is assumed the mass differences have a negligible impact on kinematics -- not favouring some mass states' formation over others'. $U_{\alpha k}$ is the unitary PMNS matrix, and the normalisation conditions are $\langle \nu_k | \nu_j \rangle = \delta_{kj}$, so that $\langle \nu_\alpha | \nu_\beta \rangle = \delta_{\alpha \beta}$. In a vacuum, it is the mass states that are eigenstates of the free Hamiltonian ($\hat{H}_0$), and so whose evolution can be computed
\begin{equation}
    \partial^{\mu} |\nu_k (x)\rangle = \hat{P}^{\mu}_0 |\nu_k (x)\rangle = p^{\mu}_k |\nu_k (x)\rangle,
    \label{1a2}
\end{equation}
where $\hat{P}^{\mu}_0$ is the (free) spacetime translation operator ($\hat{P}^{0}_0 = \hat{H}_0$) and $p^{\mu}_k$ is the four-momentum of mass state $k$. Assuming plane-wave solutions, this is solved with
\begin{equation}
    |\nu_k (x)\rangle = e^{-i p_k \cdot x} |\nu_k\rangle,
    \label{1a3}
\end{equation}
where $|\nu_k\rangle \equiv |\nu_k (0)\rangle$, which can be further simplified via the ultra-relativistic approximation
\begin{equation}
    p_k \cdot x \approx \frac{m_k^2}{2E} L,
    \label{1a3b}
\end{equation}
where $m_k$ is the mass of $\nu_k$, $E$ its energy, and $L$ is the distance travelled. One can thus arrive at the familiar vacuum transition/survival probability by defining $P_{\nu_\alpha \rightarrow \nu_{\beta}} (L, E) \equiv \big| \langle \nu_{\beta} (L) |\nu_{\alpha}\rangle \big|^2$, so that
\begin{equation}
    P_{\nu_\alpha \rightarrow \nu_{\beta}}(L, E) = \sum_{k,j} U_{\alpha k}^* U_{\beta k} U_{\alpha j} U_{\beta j}^* \text{exp}\left(-i \frac{\Delta m_{kj}^2 L}{2 E}\right),
    \label{1a4}
\end{equation}
which is governed by the mass differences $\Delta m_{kj}^2 \equiv m_k^2 - m_j^2$ and the PMNS matrix.

When taking into account matter effects, the process is not so straightforward. Recall that previously, only the free Hamiltonian was used for spacetime translations $\hat{P}^{\mu}_0$ (eqn \ref{1a2}). Strictly speaking, one should use the full Hamiltonian, which contains the CC interaction terms
\begin{equation}
    \begin{split}
        \partial^{\mu} |\nu_\alpha(x)\rangle & = \hat{P}^{\mu} |\nu_\alpha(x)\rangle \\
        & = \left(\hat{P}^{\mu}_0  + \hat{P}^{\mu}_I\right)|\nu_\alpha(x)\rangle \\
        & \approx \left(\hat{P}^{\mu}_0  + \big\langle \hat{H}^{\text{eff}}_I\big\rangle \right)|\nu_\alpha(x)\rangle,
    \end{split}
    \label{1a5}
\end{equation}
where $\big\langle \hat{H}^{\text{eff}}_I\big\rangle$ is the average effective interaction Hamiltonian, caused by coherent forward elastic scattering in matter. Incoherent scattering in matter is exceedingly unlikely ($\sigma \sim G_F^2 s$), so one need only consider coherent elastic scattering \cite{giunti2007fundamentals}. These are described by the effective Fermi theory terms
\begin{equation}
    \begin{split}
        \hat{H}_{CC}^{\text{eff}} = \frac{G_F}{\sqrt{2}} & \left[\overline{\nu}_{e} \gamma^\mu \left( 1 - \gamma^5\right) e\right] \left[\overline{e} \gamma_\mu \left( 1 - \gamma^5\right) \nu_{e}\right], \\
        \hat{H}_{NC}^{\text{eff}} = \frac{G_F}{\sqrt{2}} & \sum_{\alpha = e, \mu, \tau} \left[\overline{\nu}_\alpha \gamma^\mu \left( 1 - \gamma^5\right) \nu_\alpha\right] \\
        \times & \sum_{\psi = e, p, n} \left[\overline{\psi} \gamma_\mu \left( g^{\psi}_V - g^{\psi}_A \gamma^5\right) \psi\right],
    \end{split}
    \label{1a6}
\end{equation}
where $G_F$ is Fermi's constant, $g^{\psi}_V$ and $g^{\psi}_A$ are the vector and axial components of the coupling constants for the associated fermion $\psi$, and all the other symbols have their usual meaning. Averaging over the particles in the matter medium, one can show that \cite{giunti2007fundamentals}
\begin{equation}
    \begin{split}
        & \big\langle \hat{H}^{\text{eff}}_I\big\rangle |\nu_\alpha\rangle = V_{\alpha} |\nu_\alpha\rangle, \\
        & V_\alpha = \sqrt{2} G_F \left( \pm N_e \delta_{\alpha e} - \frac{1}{2} N_n\right),
    \end{split}
    \label{1a7}
\end{equation}
where $N_e$ and $N_n$ are the electron and neutron densities in matter respectively. The $\pm$ is $+$ for neutrinos and $-$ for antineutrinos \cite{giunti2007fundamentals}. Now, notice that the interaction Hamiltonian acts on the flavour eigenstates, while the free spacetime translation operator acts on the mass eigenstates, creating a complicated differential equation. To deal with this, the transition amplitude $\psi_{\alpha \beta}(x) = \langle\nu_\beta|\nu_\alpha (x)\rangle$ is used, so that each part of the translation operator may act on its corresponding eigenstates. The neutron density term can be factored out since it affects all flavours equally, along with other common factors, and one arrives at the ODE \cite{giunti2007fundamentals}
\begin{equation}
    i \frac{d}{dx} \boldsymbol{\Psi}_\alpha = \frac{1}{2E} H_F \boldsymbol{\Psi}_\alpha,
    \label{1a8}
\end{equation}
where $x$ is now the spacial coordinate in the direction of propagation, and
\begin{equation}
    \boldsymbol{\Psi}_\alpha =  \begin{pmatrix}
                                    \psi_{\alpha e} (x) \\
                                    \psi_{\alpha \mu} (x) \\
                                    \psi_{\alpha \tau} (x)
                                \end{pmatrix}, \;\;
    H_F = U \mathbb{M}^2 U^{\dagger} + \mathbb{A},
    \label{1a9}
\end{equation}
\begin{equation}
    \mathbb{M}^2 =  \begin{pmatrix}
                        0 & 0 & 0 \\
                        0 & \Delta m_{21}^2 & 0 \\
                        0 & 0 & \Delta m_{31}^2
                    \end{pmatrix}, \;\;
    \mathbb{A} =    \begin{pmatrix}
                        A_{CC} & 0 & 0 \\
                        0 & 0 & 0 \\
                        0 & 0 & 0
                    \end{pmatrix},
    \label{1a10}
\end{equation}
\begin{equation}
    A_{CC} = \pm 2 \sqrt{2} E G_F N_e.
    \label{1a11}
\end{equation}
Notice again that it is the mass differences that come into play, and the difference with the vacuum ODE (eqn \ref{1a2}) is isolated to the $\mathbb{A}$ matrix, which vanishes for $N_e = 0$. 


\subsection{Extant Solutions and Approximations}

There are many ways to approach this problem, which have already been written about extensively. A sample is shown here for context. First, the mass hierarchy can be exploited to separate out the contributions of each mass difference, depending on one's setup. This can ``freeze out" one of the three transition amplitudes, and if one assumes a constant matter density, can be reduced to effective two-neutrino mixing again. These expressions are very helpful to gain a qualitative understanding of the processes, such as resonances for specific values of $A_{CC}$ that lead to a strong MSW effect. See \cite{giunti2007fundamentals} or \cite{zaglauer1988mixing} for more details.

However, if the background matter density varies, or one mass difference does not totally dominate, one must keep track of all the transition amplitudes. \cite{zaglauer1988mixing} and \cite{kneller2009three} use diagonalisation to compare with the vacuum case and determine the effective mass difference and mixing angles in matter. This is the standard approach, including numerical techniques: numerically diagonalise at each iteration of the evolution equation \cite{huber2007globes, kopp2008efficient}. Meanwhile, \cite{barger1980matter} uses Lagrange's formula to determine the evolution operator
\begin{equation}
    U(L) = \sum_n \left[\prod_{m \neq n} \frac{H_F - \mathcal{E}_m \boldsymbol{1}}{\mathcal{E}_n - \mathcal{E}_m}\right] \text{exp}\left(-i \frac{\mathcal{E}_n^2 L}{2E}\right),
    \label{1b1}
\end{equation}
where $\mathcal{E}_n$ are the three eigenvalues of $H_F$, with rather involved expressions provided for constant matter density. The expression is written here explicitly due its similarity with what will be shown later on. For its part, \cite{ohlsson2000neutrino} uses the Cayley-Hamilton theorem to decompose the evolution operator into a linear combination of second order polynomials of mixing matrices, with analytic expressions for these matrices as well as their coefficients. Recently, \cite{denton2019eigenvalues} used the eigenvector-eigenvalue identity to derive relatively simple formulae for the effective mixing angles and CP violating phase, along with perturbative approximations of these and mass differences, which can be used in the vacuum expressions as normal. A summary of many of these exact and approximate techniques can be found in \cite{barenboim2019neutrino}, along with very useful accuracy and speed comparisons in the context of long baseline $\barparen{\nu}_e$ appearance experiments. \cite{fong2023analytic} provides an elegant generalisation to ($3 + N$) neutrino flavours, and a generic matter potential that can include non-standard interactions (NSI). The effects of the sterile neutrinos and NSI are studied both together and independently. These last two papers also make their codes available via public GitHub repositories, referenced therein.

Lastly, two recent papers \cite{shafaq2020effect, bustamante2019nuoscprobexact} compute a general form for the evolution operator (assuming constant matter density) in terms of a Gell-Mann basis and structure constants, using methods from \cite{ohlsson2000neutrino}. The first paper then derives perturbative expansions for particular electron density profiles in the Earth, while the second formulates a general method to compute the oscillation probability of any general time independent Hamiltonian for two or three active neutrino flavours. The initial approach of these will be followed in the next section, with some small tweaks, before being taken in another more specific direction than the original papers. These deviations will be highlighted throughout the derivation, and some extra details added for clarity. An efficient algorithm will then be constructed, which splits the computation up into two parts, so that applications which compute oscillation probabilities for a great many neutrinos need only perform the first part once as an initialisation, saving time. Some specific examples are the provided and compared to existing numerical computations by the GLoBES package and others. The calculation and behaviour of effective parameters will then briefly be covered.


\section{Solving the Differential Equation}

\subsection{The Evolution Operator}

This subsection largely follows Bushra Shafaq and Faisal Akram's method in \cite{shafaq2020effect}. First, the traceless effective Hamiltonian $H$ is defined, since the trace acts on all flavours equally and so does not contribute to mixing
\begin{equation}
    \begin{split}
        & H \equiv H_F - \frac{1}{3} \text{tr}\left[H_F\right] \boldsymbol{1}, \\
        & \text{tr}\left[H_F\right] = \Delta m_{21}^2 + \Delta m_{31}^2 + A_{CC}.
    \end{split}
    \label{2a1}
\end{equation}
The evolution equation is thus
\begin{equation}
    i \frac{d}{dx} \boldsymbol{\Psi}_\alpha = \frac{1}{2E} H \boldsymbol{\Psi}_\alpha,
    \label{2a2}
\end{equation}
where contrary to Shafaq and Akram's paper, the $1/2E$ factor is kept separate from $H$. Assuming constant matter density, this is solved with
\begin{equation}
    \boldsymbol{\Psi}_\alpha(x) = U(x) \boldsymbol{\Psi}_\alpha(0), \;\; U(x) = \text{exp}\left(-i H \frac{x}{2E}\right),
    \label{2a3}
\end{equation}
where $U(x)$ is the evolution operator. Now, these can be decomposed using the property that the Gell-Mann matrices ($\lambda^i, \; i \in \{1, ..., 8\}$) and the identity matrix form a complete orthogonal basis for $3 \times 3$ complex matrices. $H$ is traceless, so it does not need the identity matrix
\begin{equation}
    \begin{split}
        & H = h^i \lambda^i, \; h^i = \frac{1}{2} \text{tr}\left[ H \lambda^i\right], \\
        & U(x) = u_0 \boldsymbol{1} + i u_i \lambda^i, \; u_0 = \frac{1}{3} \text{tr} \left[U(x)\right], \; u_i = \frac{1}{2i} \text{tr} \left[U(x) \lambda^i\right],
    \end{split}
    \label{2a4}
\end{equation}
where from now on repeated dummy indices imply summation. These equations are derived from the Gell-Mann matrix identities $\text{tr}\left[\lambda^i \lambda^j\right] = 2 \delta_{ij}$ and $\text{tr}\left[\lambda^i\right] = 0$.

Now, some useful general results in linear algebra will be used: for a matrix $A$, with eigenvalues $\mathcal{E}[A]_n$,
\begin{equation}
    \begin{split}
        & \text{det}\left(A\right) = \prod_n \mathcal{E}[A]_n, \;  \text{tr}\left[A\right] = \sum_n \mathcal{E}[A]_n, \\
        & \text{and if $B = f(A)$ and $f$ is a holomorphic function, } \\
        & \mathcal{E}[B]_n = f(\mathcal{E}[A]_n).
    \end{split}
    \label{2a5}
\end{equation}
Therefore, recalling $U(x) = \text{exp}\left(-i Hx\right)$ and defining $\mathcal{E}[H]_n = \mathcal{E}_n$,
\begin{equation}
    \begin{split}
        & u_0 = \frac{1}{3} \sum_{n=0}^2 \text{exp}\left(-i \frac{\mathcal{E}_n x}{2E}\right).
    \end{split}
    \label{2a6}
\end{equation}
For $u_i$, first note that from $H = h^i \lambda^i$,
\begin{equation}
    \frac{\partial U(x)}{\partial h^i} = - \frac{i t}{2 E} \lambda^i U(x),
    \label{2a7}
\end{equation}
and so using the previous identities one can show
\begin{equation}
    u_i = \frac{-i}{2} \sum_{n=0}^2 \frac{\partial \mathcal{E}_n}{\partial h^i} \text{exp}\left(-i \frac{\mathcal{E}_n x}{2E}\right).
    \label{2a8}
\end{equation}
All that is needed now are expressions for the eigenvalues $\mathcal{E}_n$ of $H$. The parametric equation of a $3 \times 3$ matrix A with eigenvalues $\lambda$ is
\begin{equation}
    \begin{split}
        \text{det} \left(A - \lambda \boldsymbol{1}\right) = & - \lambda^3 + \text{tr}(A)\lambda^2 - \frac{1}{2} \left(\text{tr}(A)^2 - \text{tr}\left(A^2\right)\right) \lambda \\
        & + \text{det}(A) \\
        = & 0,
    \end{split}
    \label{2a9}
\end{equation}
so that for the traceless $H$,
\begin{equation}
    \begin{split}
        & \mathcal{E}_n^3 - 3 a_1 \mathcal{E}_n - 2 a_0 = 0, \\
        & a_1 = \frac{1}{6} \text{tr}[H^2] = \frac{1}{3} h^i h^i, \\
        & a_0 = \frac{1}{2} \text{det}\left(H\right) = \frac{1}{3} d^{ijk} h^i h^j h^k,
    \end{split}
    \label{2a10}
\end{equation}
where $d^{ijk}$ are the symmetric structure constants of the Gell-Mann matrices
\begin{equation}
    \begin{split}
        & \{\lambda^i, \lambda^j\} = \frac{4}{3} \delta_{ij} \boldsymbol{1} + 2 d^{ijk} \lambda^k, \\
        & d^{ijk} = \frac{1}{4} \text{tr} \left(\lambda^i \{\lambda^j, \lambda^k\}\right).
    \end{split}
    \label{2a11}
\end{equation}
The last relation between the determinant and structure constants in (eqn \ref{2a10}) can be derived by first multiplying the structure constant definition (eqn \ref{2a11}) (second equation) by $h^i h^j h^k$ (and summing over these indices as normal), to find
\begin{equation}
    \text{tr}\left(H^3\right) = 2 d^{ijk} h^i h^j h^k.
    \label{2a12}
\end{equation}
Then from the definition of the determinant of a $3 \times 3$ matrix
\begin{equation}
    \text{det}(H) = \frac{1}{3!} h^i h^j h^k \epsilon_{a_1 a_2 a_3} \epsilon_{b_1 b_2 b_3} \lambda^i_{a_1 b_1} \lambda^j_{a_2 b_2} \lambda^k_{a_3 b_3},
    \label{2a13}
\end{equation}
the Levi-Civita identity $\epsilon^{a_1 a_2 a_3} \epsilon_{b_1 b_2 b_3} = 3! \delta^{[a_1}_{b_1} \delta^{a_2}_{b_2} \delta^{a_3]}_{b_3}$ (the index position is irrelevent here), and recalling that the Gell-Mann matrices are traceless ($\lambda^i_{aa} = 0$), one can find
\begin{equation}
    \text{det}(H) = \frac{1}{3} \text{tr}\left(H^3\right),
    \label{2a14}
\end{equation}
and thus
\begin{equation}
    \text{det}(H) = \frac{2}{3} d^{ijk} h^i h^j h^k.
    \label{2a15}
\end{equation}
Meanwhile, taking the derivative of the parametric equation (eqn \ref{2a10}) w.r.t $h^i$ gives the needed expression
\begin{equation}
    \frac{\partial \mathcal{E}_n}{\partial h^i} = \frac{2}{3} \frac{h^i \mathcal{E}_n + d^{ijk} h^j h^k}{ \mathcal{E}_n^2 - a_1},
    \label{2a16}
\end{equation}
while different solutions to the parametric equation are used here compared to the original paper
\begin{equation}
    \mathcal{E}_n = 2\sqrt{a_1} \text{cos} \left[ \frac{1}{3} \text{cos}^{-1}\left(\frac{ a_0}{a_1^{3/2}}\right) - \frac{2\pi n}{3}\right], \;\; n \in \{0, 1, 2\}.
    \label{2a17}
\end{equation}
These are the solutions to a depressed cubic equation for real solutions, which must be real since $H$ is Hermitian (one can also check that $a_0, a_1 \in \mathbb{R}$, and $\frac{a_0^2}{4} + \frac{a_1^3}{27} < 0$, which imply the solutions are real). The evolution operator is then
\begin{equation}
    \mathcal{U}(x) = \frac{1}{3} \sum_{n=0}^2 \left(1+ \frac{\mathcal{E}_n H + Y}{\mathcal{E}_n^2 - a_1}\right) \text{exp}\left(-i \frac{\mathcal{E}_n x}{2E}\right),
    \label{2a18}
\end{equation}
with
\begin{equation}
    Y \equiv d^{ijk} h^i h^j \lambda^k = H^2 - 2 a_1 \boldsymbol{1},
    \label{2a19}
\end{equation}
which can be shown from the first equation of (eqn \ref{2a11}) and multiplied by $h^i h^j$ (summing over indices). This last relation (eqn \ref{2a19}) was not in Bushra Shafaq and Faisal Akram's paper \cite{shafaq2020effect}.

Finally, assuming a (anti)neutrino is produced in a pure flavour state $\psi_{\alpha \beta}(0) = \delta_{\alpha \beta}$ and recalling $P_{\nu_\alpha \rightarrow \nu_\beta}(x) = \big|\psi_{\alpha \beta}(x)\big|^2$, one therefore has the transition probability
\begin{equation}
    \begin{split}
        P_{\nu_\alpha \rightarrow \nu_\beta}(L, E) = & \sum_{n, m} \left(X_n\right)_{\beta\alpha} \left(X_m\right)_{\beta\alpha}^* \text{exp}\left[-i \frac{\left(\mathcal{E}_n - \mathcal{E}_m\right) L}{2E}\right], \\
        X_n = & \frac{1}{3} \left(\boldsymbol{1} + \frac{\mathcal{E}_n H + Y}{\mathcal{E}_n^2 - a_1}\right),
    \end{split}
    \label{2a20}
\end{equation}
where $x = L$ is the propagation length, as usual. This equation is of course of the same form as the vacuum case, but writing out the effective mass differences and mixing angles is saved for a later section.


\subsection{Details and Simplifications in Vacuum}

The rest of this paper departs from \cite{shafaq2020effect}, and is entirely original work. Notice that variable quantities such as $L$ and $E$ only appear in the last expression (eqn \ref{2a20}), except for where $E$ and $N_e$ enter into $A_{CC}$ at the beginning. Because of the structure of $H$ in terms of $A_{CC}$, it will turn out that most calculations can be performed with vacuum settings ($A_{CC} = 0$), and small modifications added later to take into account matter effects (see the next section). Therefore, here we take a look at the details assuming a vacuum first, where all the associated quantities will be marked with a tilde for clarity $H_F = \tilde{H}_F + \mathbb{A}$.

Here $\tilde{H}_F$ is simply $\tilde{H}_F = U \mathbb{M} U^{\dagger}$, and from the cyclic nature of the trace $\text{tr}[\tilde{H}_F] = \text{tr}[\mathbb{M}]$, so that
\begin{equation}
    \begin{split}
        & \tilde{H} = U \mathbb{M} U^{\dagger} - \frac{1}{3} \text{tr}[\mathbb{M}] \boldsymbol{1}, \\
        & \text{tr}[\mathbb{M}] = \Delta m_{21}^2 + \Delta m_{31}^2,
    \end{split}
    \label{2b1}
\end{equation}
and thus the components are explicitly given by
\begin{equation}
    \tilde{H}_{\alpha \beta} = \sum_{f=2,3} \Delta m_{f1}^2 \left(U_{\alpha f} U_{\beta f}^* - \frac{1}{3} \delta_{\alpha\beta} \right).
    \label{2b2}
\end{equation}
Now, $\tilde{a}_1$ and $\tilde{a}_0$ can be computed from $\tilde{h}^i$ and $d^{ijk}$, but it is easier to use the definitions $\tilde{a}_1 = \frac{1}{6} \text{tr}\left[\tilde{H}^2\right]$ and $\tilde{a}_0 = \frac{1}{2} \text{det}\left(\tilde{H}\right)$. For $\tilde{a}_1$ it is straightforward to show, using (eqn \ref{2b1})
\begin{equation}
    \tilde{a}_1 = \frac{1}{9} \left[ (\Delta m_{21}^2)^2 + (\Delta m_{31}^2)^2 - \Delta m_{21}^2 \Delta m_{31}^2\right],
    \label{2b3}
\end{equation}
while for $\tilde{a}_0$, the formula (eqn \ref{2a9}) for $\text{det}\left(A - \lambda \boldsymbol{1}\right)$ can be reused, with $A = U \mathbb{M} U^{\dagger}$ and $\lambda = \frac{1}{3} \text{tr}[\mathbb{M}]$, so that
\begin{equation}
    \begin{split}
        \tilde{a}_0 = & \frac{1}{27} \left[(\Delta m_{21}^2)^3 +(\Delta m_{31}^2)^3\right] \\
        & - \frac{1}{18} \left[(\Delta m_{21}^2)^2 \Delta m_{31}^2 + \Delta m_{21}^2 (\Delta m_{31}^2)^2\right],
    \end{split}
    \label{2b4}
\end{equation}
where use was made of $\text{det}\left(U \mathbb{M} U^{\dagger}\right) = \text{det}(U) \text{det}(\mathbb{M}) \text{det}(U^{\dagger})$ and $\text{det}(\mathbb{M}) = 0$. Lastly, one can show that
\begin{equation}
    \tilde{Y}_{\alpha\beta} = \frac{1}{3} \sum_{f=1}^3 \left(\Delta m^2_{f1}\right)^2 \left( U_{\alpha f} U_{\beta f}^* - \frac{1}{3} \delta_{\alpha \beta}\right),
    \label{2b5}
\end{equation}
where $\left(\Delta m^2_{11}\right)^2 \equiv 2 \Delta m^2_{21} \Delta m^2_{31}$ is defined for compactness. So for a vacuum, these quantities can all be substituted in to compute $\mathcal{E}_n$ (eqn \ref{2a17}), $X_n$ and $P_{\nu_\alpha \rightarrow \nu_\beta}(L)$ (eqn \ref{2a20}) directly. Notice also that the eigenvalues $\mathcal{E}_n$ only depend on the mass differences here, as one would expect.


\subsection{Adding Matter Effects}

The values calculated above must be corrected for matter effects. From $H_F = \tilde{H}_F + \mathbb{A}$, the traceless matrix $H$ can be related to the vacuum one $\tilde{H}$ simply according to $A_{CC}$
\begin{equation}
    H = \tilde{H} + \frac{1}{3} A_{CC} D, \;\;\;\; D = \begin{pmatrix}
                                                            2 & 0 & 0 \\
                                                            0 & -1 & 0 \\
                                                            0 & 0 & -1
                                                        \end{pmatrix}.
    \label{2c1}
\end{equation}
Corrections to $Y$ are also easier to see in matrix notation
\begin{equation}
    Y = \tilde{Y} + \frac{1}{3} A_{CC} T + \frac{1}{9} A_{CC}^2 D,
    \label{2c1.5}
\end{equation}
\begin{equation}
    T = \begin{pmatrix}
            2 \tilde{H}_{ee} & \tilde{H}_{e\mu} & \tilde{H}_{e\tau} \\
            \tilde{H}_{e\mu}^* & 2 \tilde{H}_{\tau\tau} & -2 \tilde{H}_{\mu\tau} \\
            \tilde{H}_{e\tau}^* & -2 \tilde{H}_{\mu\tau}^* & 2 \tilde{H}_{\mu\mu}
        \end{pmatrix}.
    \label{2c2}
\end{equation}
Since only the diagonal components of $\tilde{H}$ change, from $a_1 = -\frac{1}{2} \text{tr}(H^2)$ one can find
\begin{equation}
    a_1 = \tilde{a}_1 + \frac{1}{3} \tilde{H}_{ee} A_{CC} + \frac{1}{9} A_{CC}^2,
    \label{2c3}
\end{equation}
and using the determinant definition of $a_0$, it is modified by
\begin{equation}
    \begin{split}
        a_0 = & \tilde{a}_0 + \frac{1}{6} A_{CC} \left(\tilde{H}_{ee}^2 + 2 \tilde{H}_{\mu\mu} \tilde{H}_{\tau\tau} - 2 |\tilde{H}_{\mu\tau}|^2  \right. \\
        & \left. + |\tilde{H}_{e\mu}|^2 + |\tilde{H}_{e\tau}|^2\right)
        + \frac{1}{6} A_{CC}^2 \tilde{H}_{ee} + \frac{1}{27} A_{CC}^3,
    \end{split}
    \label{2c4}
\end{equation}
which one can find is simply
\begin{equation}
    a_0 = \tilde{a}_0 + \frac{1}{2} \tilde{Y}_{ee} A_{CC} + \frac{1}{6} \tilde{H}_{ee} A_{CC}^2 + \frac{1}{27} A_{CC}^3.
    \label{2c5}
\end{equation}

Notice that since the diagonal components of $\tilde{H}$ are real, so are those of $\tilde{Y}$, and therefore $a_0$ and $a_1$ and, by extension $\mathcal{E}_n$, are always real. $X_n$ consequently always has real diagonal components, as expected. Finally, also note that $Y$ and $H$ (eqn \ref{2c1.5}, \ref{2c2}) have higher order matter corrections in their diagonal components. This means that survival probabilities appear to be more greatly affected by constant matter density than transition probabilities are.


\section{Example Algorithms}

The following section contains a few specific examples of algorithms one can implement from the method derived above. This is to gather all the relevant information in one convenient place for any reader simply wishing to apply this method.


\subsection{Electron (Anti)Neutrino Survival Probability}

What turns out to be the simplest example of how this can all be used in an algorithm is shown here. It is composed of two steps: the first performed once to compute some constant values, and the second using these values for each particular (anti)neutrino energy and/or electron density.

First one should compute the following four constant quantities, written here in terms of the mass differences and mixing angles of the standard PMNS matrix parametrisation:
\begin{equation}
    \tilde{H}_{ee} = \Delta m_{21}^2 \left(s_{12}^2 c_{13}^2 - \frac{1}{3}\right) + \Delta m_{31}^2 \left(s_{13}^2 - \frac{1}{3}\right),
    \label{2d1}
\end{equation}
\begin{equation}
    \begin{split}
        \tilde{Y}_{ee} = \frac{1}{3} \bigg[& \left(\Delta m_{21}^2\right)^2 \left(s_{12}^2 c_{13}^2 - \frac{1}{3}\right) \\
        & + \left(\Delta m_{31}^2\right)^2 \left(s_{13}^2 - \frac{1}{3}\right) \\
        & + 2 \Delta m_{21}^2 \Delta m_{31}^2 \left( c_{12}^2 c_{13}^2 - \frac{1}{3}\right)\bigg],
    \end{split}
    \label{2d2}
\end{equation}
\begin{equation}
    \begin{split}
        \tilde{a}_0 = & \frac{1}{27} \left[(\Delta m_{21}^2)^3 +(\Delta m_{31}^2)^3\right] \\
        & - \frac{1}{18} \left[(\Delta m_{21}^2)^2 \Delta m_{31}^2 + \Delta m_{21}^2 (\Delta m_{31}^2)^2\right],
    \end{split}
    \label{2d3}
\end{equation}
and
\begin{equation}
    \tilde{a}_1 = \frac{1}{9} \left[ (\Delta m_{21}^2)^2 + (\Delta m_{31}^2)^2 - \Delta m_{21}^2 \Delta m_{31}^2\right].
    \label{2d4}
\end{equation}
 
Then the next part is performed for a given electron (anti)neutrino energy $E$ and matter electron density $N_e$. First $A_{CC}$ is computed:
\begin{equation}
    A_{CC} = \pm 2 \sqrt{2} G_F E N_e,
    \label{2d5}
\end{equation}
then the constants corrected for this:
\begin{equation}
    H_{ee} = \tilde{H}_{ee} + \frac{2}{3} A_{CC},
    \label{2d6}
\end{equation}
\begin{equation}
    a_0 = \tilde{a}_0 + \frac{1}{2} \tilde{Y}_{ee} A_{CC} + \frac{1}{6} \tilde{H}_{ee} A_{CC}^2 + \frac{1}{27} A_{CC}^3,
    \label{2d7}
\end{equation}
\begin{equation}
    a_1 = \tilde{a}_1 + \frac{1}{3} \tilde{H}_{ee} A_{CC} + \frac{1}{9} A_{CC}^2,
    \label{2d8}
\end{equation}
\begin{equation}
    Y_{ee} = \tilde{Y}_{ee} + \frac{2}{3} \tilde{H}_{ee} A_{CC} + \frac{2}{9} A_{CC}^2.
    \label{2d9}
\end{equation}
These can then be substituted into
\begin{equation}
    \mathcal{E}_n = 2\sqrt{a_1} \text{cos} \left[ \frac{1}{3} \text{cos}^{-1}\left(\frac{ a_0}{a_1^{3/2}}\right) - \frac{2\pi n}{3}\right], \;\; n \in \{0, 1, 2\}.
    \label{2d10}
\end{equation}
\begin{equation}
    \left(X_n\right)_{ee} = \frac{1}{3} \left( 1 + \frac{\mathcal{E}_n H_{ee} + Y_{ee}}{ \mathcal{E}_n^2 - a_1}\right),
    \label{2d11}
\end{equation}
so that finally
\begin{equation}
    P_{\overset{\textbf{\fontsize{2pt}{2pt}\selectfont(---)}}{\nu_e} \rightarrow \overset{\textbf{\fontsize{2pt}{2pt}\selectfont(---)}}{\nu_e}} = 1 - 4 \sum_{n>m} (X_n)_{ee}(X_m)_{ee} \text{sin}^2\left(\left(\mathcal{E}_n - \mathcal{E}_m\right) \frac{L}{4E}\right).
    \label{2d12}
\end{equation}
The fact that the components $(X_n)_{ee}$ are real was used to derive the more compact formula (eqn \ref{2d12}). Notice that neither $\theta_{23}$, nor $\delta_{13}$ enter this calculation, no matter the electron density. The (anti)electron neutrino survival probability is thus independent of these for constant densities.


\subsection{Muon (Anti)Neutrino Survival Probability}

One can follow the exact same method for muon neutrinos, with a couple small tweaks. In the first step, the two following real numbers must also be computed
\begin{equation}
    \begin{split}
        \tilde{H}_{\mu\mu} = \Delta m_{21}^2 \bigg( & c_{12}^2 c_{23}^2 + s_{12}^2 s_{13}^2 s_{23}^2 - 2 s_{12} s_{13} s_{23} c_{12} c_{23} \text{cos}\delta \\
        & - \frac{1}{3}\bigg) + \Delta m_{31}^2 \bigg( s_{23}^2 c_{13}^2 - \frac{1}{3}\bigg),
    \end{split}
    \label{2f1}
\end{equation}
\begin{equation}
    \begin{split}
        \tilde{Y}_{\mu\mu} = \frac{1}{3} \bigg[& \left(\Delta m_{21}^2\right)^2 \bigg( c_{12}^2 c_{23}^2 + s_{12}^2 s_{13}^2 s_{23}^2 \\
        & - 2 s_{12} s_{13} s_{23} c_{12} c_{23} \text{cos}\delta - \frac{1}{3}\bigg) \\
        & + \left(\Delta m_{31}^2\right)^2 \bigg(s_{23}^2 c_{13}^2 - \frac{1}{3}\bigg) \\
        & + 2 \Delta m_{21}^2 \Delta m_{31}^2 \bigg( s_{12}^2 c_{23}^2 + s_{13}^2 s_{23}^2 c_{12}^2 \\
        & + 2 s_{12} s_{13} s_{23} c_{12} c_{23} \text{cos}\delta - \frac{1}{3}\bigg)\bigg],
    \end{split}
    \label{2f2}
\end{equation}
on top of the four constants $\tilde{a}_1$, $\tilde{a}_0$, $\tilde{H}_{ee}$ and $\tilde{Y}_{ee}$ which are needed for all flavour oscillations. Now, only $a_1$ and $a_0$ of these must be corrected with $A_{CC}$ (eqn \ref{2d7}, \ref{2d8}), along with
\begin{equation}
    H_{\mu\mu} = \tilde{H}_{\mu\mu} - \frac{1}{3} A_{CC},
    \label{2f3}
\end{equation}
\begin{equation}
    Y_{\mu\mu} = \tilde{Y}_{\mu\mu} - \frac{2}{3} \left(\tilde{H}_{ee} + \tilde{H}_{\mu\mu}\right) A_{CC} - \frac{1}{9} A_{CC}^2,
    \label{2f4}
\end{equation}
where $\tilde{H}_{\tau\tau} = - \left(\tilde{H}_{ee} + \tilde{H}_{\mu\mu}\right)$ was used to arrive at (eqn \ref{2f4}). Then the eigenvalues $\mathcal{E}_n$ are computed exactly as before (eqn \ref{2d10}), and all substituted into the formulae
\begin{equation}
    \left(X_n\right)_{\mu\mu} = \frac{1}{3} \left( 1 + \frac{\mathcal{E}_n H_{\mu\mu} + Y_{\mu\mu}}{ \mathcal{E}_n^2 - a_1}\right),
    \label{2f5}
\end{equation}
\begin{equation}
    P_{\overset{\textbf{\fontsize{2pt}{2pt}\selectfont(---)}}{\nu_\mu} \rightarrow \overset{\textbf{\fontsize{2pt}{2pt}\selectfont(---)}}{\nu_\mu}} = 1 - 4 \sum_{n>m} (X_n)_{\mu}(X_m)_{\mu} \text{sin}^2\left(\left(\mathcal{E}_n - \mathcal{E}_m\right) \frac{L}{4E}\right).
    \label{2f6}
\end{equation}


\subsection{Electron (Anti)Neutrino Appearance}

Another more complex example is muon (anti)neutrino to electron (anti)neutrino transition probability. Just as before, first compute the vacuum values $\tilde{a}_1$, $\tilde{a}_0$, $\tilde{H}_{ee}$ and $\tilde{Y}_{ee}$.
However, in this case since we are dealing with transition probabilities, one must additionally compute the two complex numbers
\begin{equation}
    \begin{split}
        \tilde{H}_{e\mu} = & \Delta m_{21}^2 s_{12} c_{13} \left(c_{12} c_{23} - s_{12} s_{23} s_{13} e^{-i \delta}\right) \\
        & + \Delta m_{31}^2 s_{13} s_{23} c_{13} e^{-i \delta},
    \end{split}
    \label{2e1}
\end{equation}
and
\begin{equation}
    \begin{split}
        \tilde{Y}_{e\mu} = \frac{1}{3} \bigg[ & \left(\Delta m_{21}^2\right)^2 s_{12}c_{13} \left( c_{12}c_{23} - s_{12}s_{13}s_{23} e^{-i\delta} \right) \\
        & + \left(\Delta m_{31}^2\right)^2 s_{13}s_{23}c_{13} e^{-i\delta} \\
        & - 2 \Delta m_{21}^2 \Delta m_{31}^2 c_{12}c_{13} \left( s_{12}c_{23} + s_{13}s_{23}c_{12} e^{-i\delta}\right) \bigg].
    \end{split}
    \label{2e2}
\end{equation}
There are thus overall eight real quantities $\tilde{H}_{ee}$, $\tilde{a}_0$, $\tilde{a}_1$, $\tilde{Y}_{ee}$, $\Re[\tilde{Y}_{e\mu}]$, $\Im[\tilde{Y}_{e\mu}]$, $\Re[\tilde{H}_{e\mu}]$ and $\Im[\tilde{H}_{e\mu}]$ that are pre-computed and passed on to be corrected at run time for matter effects.

With a given (anti)neutrino energy and electron density, $A_{CC}$ is computed as in (eqn \ref{2d5}), then $a_0$, $a_1$ and $\mathcal{E}_n$ are corrected and computed respectively just as previously (eqn \ref{2d7}, \ref{2d8}, \ref{2d10}). These are then substituted into
\begin{equation}
    \left(X_n\right)_{e\mu} = \frac{1}{3} \frac{\left(\mathcal{E}_n + \frac{1}{3}A_{CC}\right) \tilde{H}_{e\mu} + \tilde{Y}_{e\mu}}{\mathcal{E}_n^2 - a_1},
    \label{2e4}
\end{equation}
so that finally
\begin{equation}
    \begin{split}
         P_{\overset{\textbf{\fontsize{2pt}{2pt}\selectfont(---)}}{\nu_\mu} \rightarrow \overset{\textbf{\fontsize{2pt}{2pt}\selectfont(---)}}{\nu_e}}(L) = - 4 \sum_{n>m} & \left(R_n R_m  + I_n I_m\right) \text{sin}^2\left(\left(\mathcal{E}_n - \mathcal{E}_m\right)\frac{L}{4E}\right) \\
         \pm 2 \sum_{n>m} & \left(I_n R_m - R_n I_m\right) \text{sin}\left(\left(\mathcal{E}_n - \mathcal{E}_m\right)\frac{L}{2E}\right),
    \end{split}
    \label{2e5}
\end{equation}
where the $\pm$ is positive for neutrinos and negative for antineutrinos, while
\begin{equation}
    \begin{split}
        & R_n \equiv \Re\left[\left(X_n\right)_{e\mu}\right], \\
        & I_n \equiv \Im\left[\left(X_n\right)_{e\mu}\right].
    \end{split}
    \label{2e6}
\end{equation}


\section{Results}

\subsection{Speed Comparison for Example Algorithm}

In order to get a sense of the speed of this algorithm, calculations of various neutrino oscillation probabilities in constant matter density were performed by the above algorithms, written in C++. The same was then done with the widely used GLoBES package \cite{huber2007globes} - also written in C++ - and the results and processing times of the two were compared. Use was made of the \textit{glbConstantDensityProbability()} function, which computes the transition or survival probability between any two neutrino flavours for constant matter density, neutrino energy and baseline. It does this by diagonalising the Hamiltonian with various numerical or analytic methods, depending on which is fastest \cite{ohlsson2000neutrino, kopp2008efficient}. It is a fast and reliable method, but all matrix elements must be recomputed for each change in neutrino energy $E$, baseline $L$ and matter density $\rho$, while this paper's method must only re-perform part of the calculations.

\subsubsection{Method}

First, an initialisation step is performed, where the pre-computed constants above are calculated and GLoBES is initialised. This step was not timed since it need only be performed once and so will not scale with the number of calculations. However, note that the GLoBES initialisation takes longer since the package includes many more functionalities than just constant matter neutrino oscillations.

Second, for a given flavour transition, and a range of 100 neutrino energies, 100 baselines and 100 matter densities, the oscillation probabilities were computed using three different functions: the GLoBES function, a general flavour version of this algorithm (Section II), and then the version of this algorithm tailored to the specific flavour transition (Section III). The results and total computation times (CPU time as measured by the C++ standard library \textit{std::clock()} function) of these three were recorded. This process was repeated 50 times to obtain a measure of statistical uncertainty.

The ranges of neutrino energies $E$, baselines $L$ and matter densities $\rho$ were evenly spaced values in some range
\begin{equation}
    \begin{split}
        & E_{\text{min}} \leq E \leq E_{\text{max}}, \\
        & L_{\text{min}} \leq L \leq L_{\text{max}}, \\
        & \rho_{\text{min}} \leq \rho \leq \rho_{\text{max}},
    \end{split}
\end{equation}
where the minima were always fixed ($E_{\text{min}} = 0.5$ MeV, $L_{\text{min}} = 0.01$ km, $\rho_{\text{min}} = 0$ g/cm${}^3$). Therefore, for a given set of maxima ($E_{\text{max}}$, $L_{\text{max}}$, $\rho_{\text{max}}$), each function was called 50 million times ($100 \times 100 \times 100 \times 50$).

\subsubsection{Results}

As alluded to, the whole process was performed for various maximum values, to discern any $E$, $L$ or $\rho$ dependence on the results. Ten different values for each were used, according to
\begin{equation}
    \begin{split}
        & E_{\text{max}} \in \{x \; : \; E_{\text{min}} \leq x \leq 1000 \text{ MeV}\}, \\
        & L_{\text{max}} \in \{x \; : \; L_{\text{min}} \leq x \leq 1000 \text{ km}\}, \\
        & \rho_{\text{max}} \in \{x \; : \; \rho_{\text{min}} \leq x \leq 100 \text{ g/cm}^3\},
    \end{split}
\end{equation}
so that the whole method above was carried out one thousand times.

\begin{figure}[h]
    \centering
    \begin{subfigure}[b]{0.3\textwidth}
         \centering
         \includegraphics[width=\textwidth]{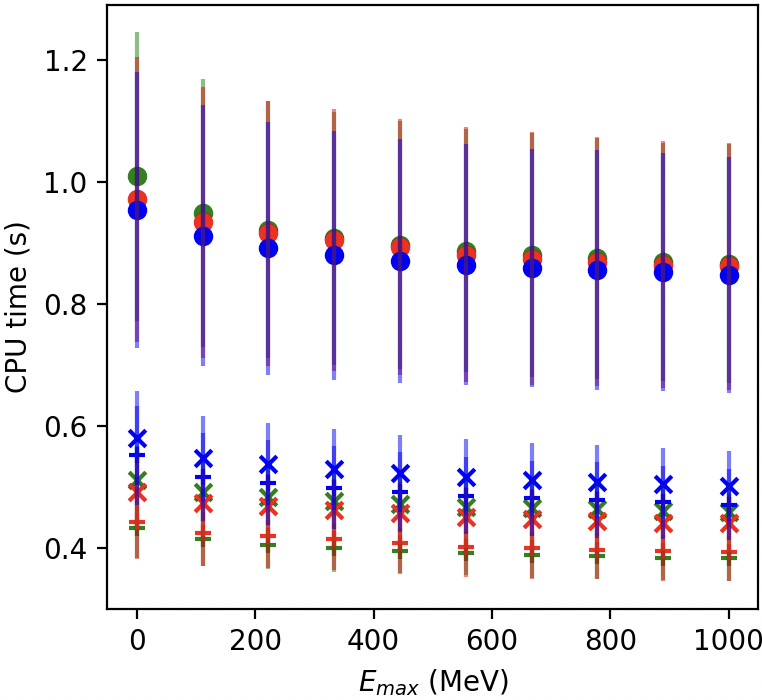}
         \caption{Energy dependence}
         \label{fig1a}
     \end{subfigure}
     \hfill
     \begin{subfigure}[b]{0.3\textwidth}
         \centering
         \includegraphics[width=\textwidth]{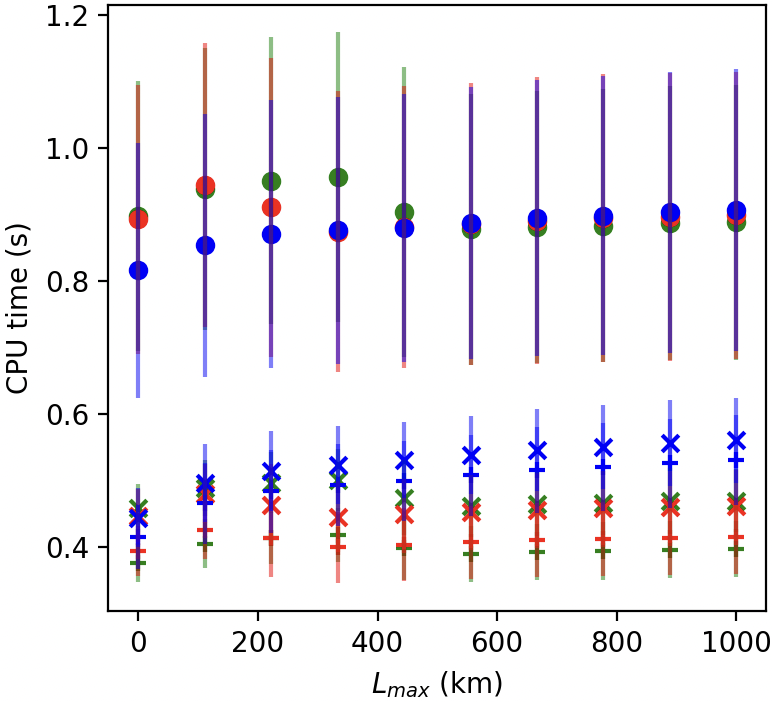}
         \caption{Baseline dependence}
         \label{fig1b}
     \end{subfigure}
     \hfill
     \begin{subfigure}[b]{0.3\textwidth}
         \centering
         \includegraphics[width=\textwidth]{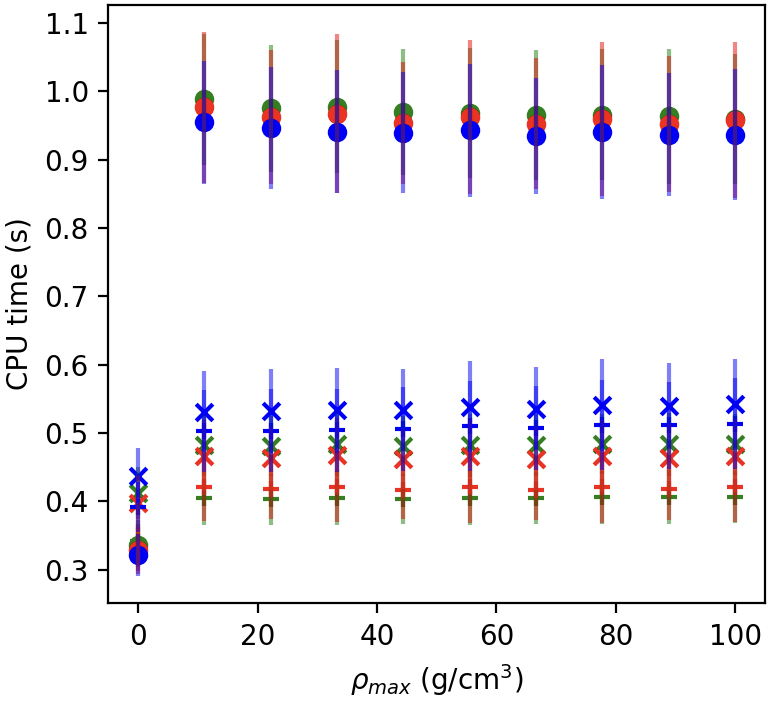}
         \caption{Matter density dependence}
         \label{fig1c}
     \end{subfigure}
     \hfill
     \begin{subfigure}[b]{0.17\textwidth}
         \centering
         \includegraphics[width=\textwidth]{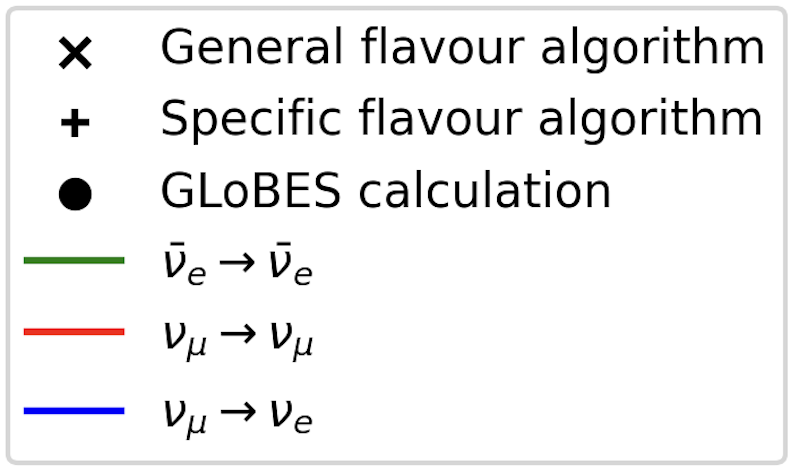}
         \caption{Legend}
         \label{fig1d}
     \end{subfigure}
    \caption{Comparison of the CPU time taken by new neutrino oscillation algorithms against the same GLoBES calculation.}
    \label{fig1}
\end{figure}

The computed probabilities were always exactly the same (having copied any unit conversion factors from the GLoBES code), so are not shown here. However, the computation times are presented in figure \ref{fig1} for three example oscillations. Dependence on the three $E_{\text{max}}$, $L_{\text{max}}$ and $\rho_{\text{max}}$ parameters is shown separately. The plot showing $E_{\text{max}}$ dependence averages over all $L_{\text{max}}$ and $\rho_{\text{max}}$ dependence, and likewise for the other two cases. Statistical error was propagated throughout, and added quadratically to a systematic error or $\sigma_{\text{sys}} = 0.01$ s, being the resolution limit of the CPU time measuring function.

For zero matter density, the GLoBES calculation appears slightly faster, though is still within one standard deviation from this paper's algorithms. Every other configuration shows these to be almost or around twice as fast as GLoBES. To be specific, if $T_{\text{GLoBES}}$ is the CPU time taken by the GLoBES function averaged over all the data above, and likewise $T_{\text{general}}$ and $T_{\text{specific}}$ are the CPU times taken by the general and specific flavour algorithms from this paper respectively,
\begin{equation}
    \frac{T_{\text{GLoBES}}}{T_{\text{general}}} = 1.82 \pm 0.34, \;\;\;\; \frac{T_{\text{GLoBES}}}{T_{\text{specific}}} = 1.98 \pm 0.36.
\end{equation}
Most of the uncertainty comes from the variability in GLoBES' computation time.

Additionally, following discussion with Peter Denton, the $\barparen{\nu}_e$ appearance algorithm was compared to other calculations described in \cite{barenboim2019neutrino}. Use was made of Peter Denton's code on github, where a fork was made (\url{https://github.com/Jamicus96/Nu-Pert-Compare}) to add this paper's algorithm in two separate cases:
\begin{itemize}
    \item The first step, or initialisation of quantities in vacuum, is performed separately, before any speed comparison (branch ``compare\_JP\_precomp''). \\
    \item The initialisation is included in the speed comparison (branch ``compare\_JP'').
\end{itemize}
It was found that the former was marginally faster than the fastest exact calculation in Peter Denton's code package ("ZS") -- on the order of 6\% faster -- while the latter slightly slower -- around 17\% slower. It was estimated that accounting for the initialisation time, this paper's algorithm becomes faster than the "ZS" algorithm after 3 loops (3 probability calculations for different neutrino energies). The differences here are small, and one must bear in mind that some approximate solutions described in \cite{barenboim2019neutrino}, and included in the code, are significantly faster.


\subsection{Effective Parameters}

\begin{figure}[h]
    \centering
    \begin{subfigure}[b]{0.46\textwidth}
        \centering
        \includegraphics[width=\textwidth]{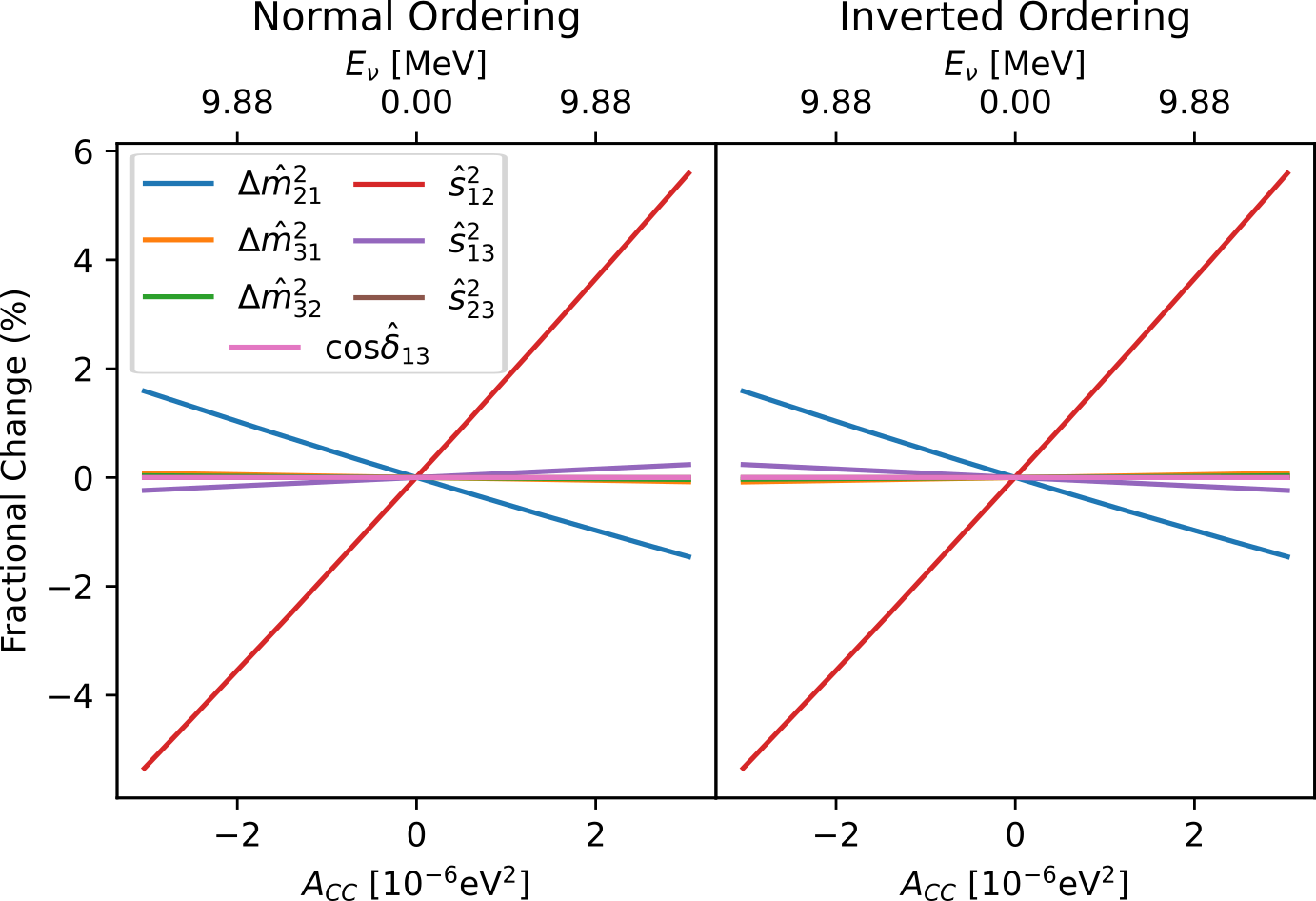}
        \caption{MeV scale (for $\rho = 2.7$g/cm${}^3$ and $\langle N/A\rangle = 0.5$).}
        \label{fig3.1}
    \end{subfigure}
    \hfill
    \begin{subfigure}[b]{0.46\textwidth}
        \centering
        \includegraphics[width=\textwidth]{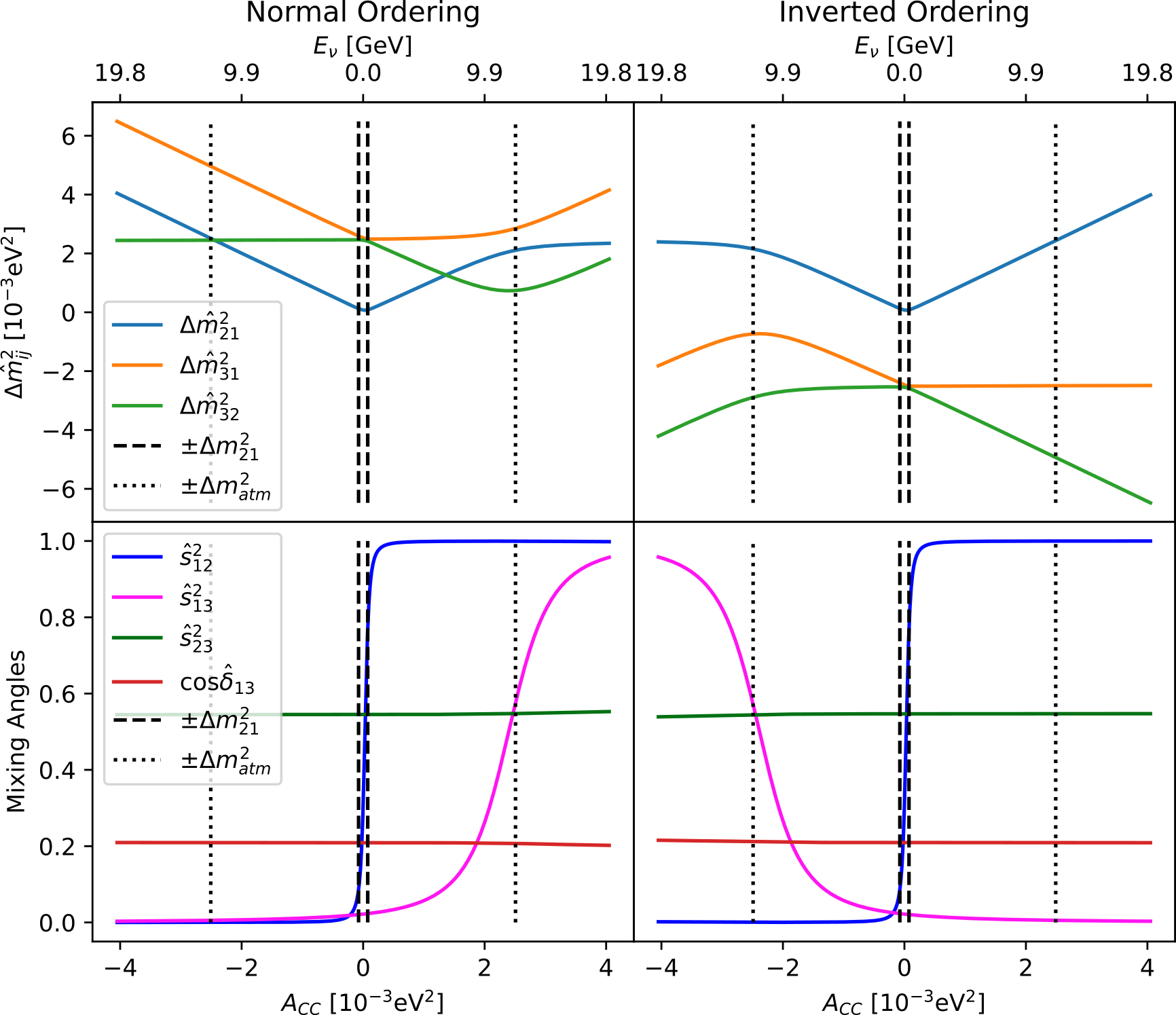}
        \caption{GeV scale (for $\rho = 2.7$g/cm${}^3$ and $\langle N/A\rangle = 0.5$).}
        \label{fig3.2}
    \end{subfigure}
    \caption{Change of the effective neutrino oscillation constants with $A_{CC}$. Positive values of $A_{CC}$ represent neutrinos, while negative values represent antineutrinos. Vertical lines denote where $A_{CC}$ is equal to the vacuum squared mass-differences, and specifically $\Delta m_{\text{atm}}^2 \equiv \frac{1}{2} \left(\Delta m_{\text{32}}^2 + \Delta m_{\text{31}}^2\right)$ is the average atmospheric oscillation mass-difference.}
\end{figure}

From the correspondence between the PMNS matrix and $X_n$ matrices found earlier (eqn \ref{2a20}), as well as that between the eigenvalue differences and the mass differences, one can find the effective parameters (including matter effects)
\begin{equation}
    \begin{split}
        & \widehat{\Delta m_{kj}^2} = \mathcal{E}_m - \mathcal{E}_n, \\
        & \widehat{U}_{\alpha k} \widehat{U}_{\beta k}^* = \left(X_n\right)_{\alpha\beta},
    \end{split}
    \label{3b1}
\end{equation}
for some relationship between $(k, j)$ and $(m, n)$ indices. To deduce this relationship, note that (\ref{3b1}) must hold for the vacuum case, so the relationship need only be shown for that simplified case. Now, the eigenvalues of the traceless matrix $\tilde{H}$ (\ref{2a1}) in the vacuum case are clearly
\begin{equation}
    \begin{split}
        & \lambda_1 = - \frac{1}{3} \left(\Delta m_{21}^2 + \Delta m_{31}^2\right), \\
        & \lambda_2 = \frac{2}{3} \Delta m_{21}^2 - \frac{1}{3} \Delta m_{31}^2, \\
        & \lambda_3 =  - \frac{1}{3} \Delta m_{21}^2 + \frac{2}{3} \Delta m_{31}^2,
    \end{split}
\end{equation}
so the vacuum eigenvalues $\tilde{\mathcal{E}}_n$ must be assigned to these in some order.
Next, looking at the definition of $\mathcal{E}_n$ (\ref{2a17}), the $\frac{1}{3} \text{cos}^{-1}(...)$ term is always between $0$ and $\frac{\pi}{3}$. This means that
\begin{equation}
    \mathcal{E}_0 \; > \; \mathcal{E}_1 \; > \; \mathcal{E}_2,
    \label{eigen_ineq}
\end{equation}
and $\mathcal{E}_0 > 0$ always hold. Therefore, the ordering of $\lambda_1$, $\lambda_2$ and $\lambda_3$ determines their relationship. This ordering depends on the mass ordering itself: for Normal Ordering (NO) $\lambda_3 > 0 > \lambda_2 > \lambda_1$, and for Inverted Ordering (IO) $\lambda_2 > \lambda_1 > 0 > \lambda_3$. Therefore, the relationship between the indices in (\ref{3b1}) is shown in table \ref{table:1}.
\begin{table}[h!]
    \centering
    \begin{tabular}{| c | c | c |}
        \hline
        (k, j) & NO (m, n) & IO (m, n) \\
        \hline
        3 & 0 & 2\\
        2 & 1 & 0\\
        1 & 2 & 1\\
        \hline
    \end{tabular}
    \caption{Index correspondence for effective parameters from equation \ref{3b1}, for Normal Ordering (NO) and Inverted Ordering (IO).}
    \label{table:1}
\end{table}
For example in normal ordering, $\widehat{\Delta m_{21}^2} = \mathcal{E}_1 - \mathcal{E}_2$ and $|\widehat{U}_{e 3}|^2 = \left(X_0\right)_{ee}$. Thus, the effective mixing angles can be evaluated in terms of these too, such as in normal ordering for example:
\begin{equation}
    \begin{split}
        & \widehat{s_{13}}^2 = \left(X_0\right)_{ee}, \\
        & \widehat{s_{12}}^2 = \frac{\left(X_1\right)_{ee}}{\widehat{c_{13}}^2}, \\
        & \widehat{s_{23}}^2 = \frac{\left(X_0\right)_{\mu\mu}}{\widehat{c_{13}}^2}, \\
        & \text{cos}\widehat{\delta_{13}} = \frac{\widehat{c_{12}}^2 \widehat{c_{23}}^2 + \widehat{s_{12}}^2 \widehat{s_{23}}^2 \widehat{s_{13}}^2 - \left(X_1\right)_{\mu\mu}}{2 \widehat{c_{12}} \widehat{c_{23}} \widehat{s_{12}} \widehat{s_{23}} \widehat{s_{13}}}.
    \end{split}
    \label{3b2}
\end{equation}
Notice that any dependence on energy or electron density in these parameters comes only from factors of $A_{CC}$. Their values can thus be plotted on a simple graph against $A_{CC} \propto E N_e$, without having to vary $E$ and $N_e$ independently. Additionally, negative values of $A_{CC}$ can be used to plot the behaviour of antineutrinos, since flipping the sign of $A_{CC}$ is effectively the only difference. See figure \ref{fig3.1} for the fractional scaling of these parameters in the MeV neutrino energy scale (for lithospheric electron density). To see large scale absolute changes, such as mass differences changing ordering, one must go to the GeV scale, as shown in figure \ref{fig3.2}. As one might expect, ``regime changes" (sudden changes in evolution of effective parameters) appear when $A_{CC}$ reaches the same scale as the mass differences. Notice also that $\widehat{\Delta m_{21}^2}$ and $\widehat{s_{12}}^2$ are the most strongly affected by matter in the MeV scale.

Furthermore, one can use these effective oscillation parameters in short and long baseline approximate formulae, just as one would with the vacuum constants. For example the long baseline approximation
\begin{equation}
    P = 1 - \frac{1}{2} \text{sin}^2 \left(2\theta_{13}\right) - \text{sin}^2 \left(2\theta_{12}\right) c_{13}^2 \text{sin}^2 \left( \frac{L \Delta m_{21}^2}{4 E} \right),
\end{equation}
becomes, for Normal Ordering,
\begin{equation}
    P = 1 - 4 \left(X_1\right)_{ee} \left(X_2\right)_{ee} - 2 \left(X_0\right)_{ee} \text{sin}^2 \left( \frac{L \left(\mathcal{E}_1 - \mathcal{E}_2\right)}{4 E} \right),
\end{equation}
which follows the full matter effect oscillation formula more closely.


\section{Conclusions}

Clearly, the derived algorithm for constant matter density transition and survival probabilities is relatively simple and efficient compared with both numerical and previous analytic solutions \cite{zaglauer1988mixing, kneller2009three, barger1980matter, huber2007globes, ohlsson2000neutrino, kopp2008efficient}.

First the fact that this is an exact solution allows one to draw useful information such as anti electron neutrino survival probability being independent of the CP-phase for constant matter densities, and the effective parameters being only dependent on $A_{CC}(E, N_e)$ - not independently on $E$ and $N_e$. 

Second, the performance of the example algorithms is noteworthy, computing roughly two times faster than the GLoBES package for non-zero matter density. Meanwhile, GLoBES is roughly as fast for zero matter density, being less than one standard deviation away. Ignoring the pre-computation of terms independent of energy and baseline, this paper's algorithm is also 6\% faster than the fast ``ZS" algorithm described above. Including this pre-computing as an initialisation, it becomes faster than ``ZS" after 3 calculations. The algorithm presented here is thus optimal for applications involving a large number of oscillation probability calculations at different energies or baselines, such as oscillation analyses.

At the very least, this presents a relatively easy to implement and fast tool to compute oscillation probabilities for approximately constant matter density profiles, which scales up well for large numbers of calculations. It does not require the use and implementation of an entire package such as GLoBES, when only a simple oscillation probability is desired.


\section{Acknowledgements}

The author would like to thank Dr. Elisabeth Falk for her invaluable advice and help in writing and preparing this paper for publication. The author is funded by STFC at the University of Sussex, but received no specific or additional grants for the work presented here.


\bibliography{Matter_effect_refs} 


\end{document}